\documentclass[prb,showpacs,showkeys]{revtex4}

\usepackage{graphicx}
\usepackage{dcolumn}
\usepackage{bm}

\begin{document}

\preprint{APS/123-QED}

\title{Analytical Tight-binding Approach for Ballistic Transport\\ through Armchair Graphene Ribbons:\\ Exact Solutions for Propagation through\\ Step-like and Barrier-like Potentials}

\author{Yu. Klymenko}
\email{yurkl@ikd.kiev.ua}
\affiliation{%
Space Research Institute of NAS and NSA of Ukraine, Kyiv, 03187, Ukraine}%

\author{O. Shevtsov}
\email{shevtsov@univ.kiev.ua}
\affiliation{
National Taras Shevchenko University of Kyiv, 03022, Kyiv, Ukraine}%

\date{\today}

\begin{abstract}

Based on a tight-binding approximation, we present analytical solutions for the wavefunction and propagation velocity of an electron in armchair graphene ribbons. The derived expressions are used for computing the transmission coefficients through step-like and barrier-like potentials. Our analytical solutions predict a new kind of transmission resonances for one-mode propagation in semiconducting ribbons. Contrary to the Klein paradox in graphene, this approach shows that backscattering for gapless mode is possible. In consistence with a higher order $\bf{ k\cdot p}$ method, the backscattering probabilities vary with the square of the applied  potential in the low-energy limit. We also demonstrate that
gapless-mode propagation through a potential step in armchair ribbons can be described by the same
through-step relation as that for an undimerized 1D chain of identical atoms.
\end{abstract}

\maketitle

\section{Introduction}\label{1}

Recently graphene sheet, a monolayer of covalent-bond carbon
atoms forming a dense honeycomb crystal, has been obtained
experimentally \cite{Geim,Novoselov}. Even at room temperature
submicron graphene structures act as a high-mobility electron or
hole conductors \cite{Geim}.
This phenomenon has brought to life a promising field of carbon-based nanoelectronics,
where graphene ribbons (GRs) could be used as connections in
nanodevices.

The presence of armchair- and zigzag-shaped edges in graphene has
strong implications for the spectrum of $\pi$-electrons
\cite{Brey,Fujita,Nakada,Waka,Malysheva} and drastically changes
the conducting properties of GRs. Especially, a zigzag edge
provides localized edge states close to the Fermi level ($E_F=0$),
which leads to the metallic type of conductivity. In contrast,
any localized state does not appear in an armchair GR. An armchair GR
can be easily made to be either metallic or semiconducting by
controlling the width of the current channel.

Most of the intriguing electronic properties of GRs were first predicted by the tight-binding model
\cite{Fujita,Nakada,Waka,Ezawa,Robinson,Shomerus,Peres}. These features are also well reproduced in the $\bf{ k\cdot p}$ method \cite{Ando2,Ando} based on the decomposition of linear Schr\"{o}dinger equations in the vicinity of zero-energy point. The commonly used $\bf{ k\cdot p}$ equation is a
two-dimensional analog of the relativistic Dirac equation \cite{Geim,Brey,Ando,Tworzydlo}, which is a certain approximation of general
tight-binding Schr\"{o}dinger equations. The continuum
Dirac description of the electronic states near $E=E_F$ has been shown  to be quite
accurate by comparing with a numerical solution in the
nearest-neighbor tight-binding model \cite{Brey,Tworzydlo}.
Particularly, in the
framework of the Dirac approach, perfect
electron transmission is predicted  through any high and wide potential
barriers, known as the Klein paradox in graphene \cite{Kats}.

Recent analytical studies based on the relativistic Dirac equation have
considered transmission through barriers \cite{Kats}, graphene wells
\cite{Waka2,Kats2}, double-barriers
\cite{Pereira}, quantum dots \cite{Silvestrov}, superlattices \cite{Barbier},  and $n-p$ junctions
\cite{Cheianov}. However, transport properties of these structures
in the tight-binding model has not been analyzed systematically yet. Considerable attention has been paid to
the effects of conductance quantization in GRs in the absence of scattering, where the
quantization steps clearly indicate the number of propagating states crossing the Fermi level \cite{Peres}.

In this paper, we develop an exact model of electron transport through armchair GRs based on the  nearest-neighbor tight-binding approximation. (In contrast to the zigzag GRs, the armchair GR spectrum does not have a special band of edge states complicating the description.) We obtain simple analytical expressions
for normalized wavefunctions and propagation velocity of an electron wave in armchair ribbons. No similar explicit solutions have been obtained for armchair GRs yet. In a tight-binding study by Zheng \cite{Zheng} of the electronic structure in armchair ribbons, the derived analytical form of wavefunction has not been used for describing charge transport in GRs.

On the basis of the analytical solutions for the wavefunction and propagation velocity of an electron in an infinite armchair GR, we describe the scattering problem through step-like and barrier-like profiles of site energy along the ribbon in closed form. Solving the corresponding equations for the scattering amplitudes, we obtain exact transmission coefficients in the atomistic (tight-binding) model analytically. For the energy region close to the Fermi level we compare our analytical expressions with their analogs obtained in the continuous (Dirac) model \cite{Kats}. Our general conclusion is that the validity of the relativistic approach based on the standard $\bf{ k\cdot p}$ equation \cite{Ando} is overestimated for  the description of charge transport in armchair GRs. Specifically, for one-mode propagation in semimetal GRs, the deviation of transmission coefficients $T$ from the unity near the zero-energy point is  proportional to the square of the step (barrier) magnitude, which contradicts the known Klein paradox in graphene \cite{Kats}. It is worth noting that similar deviations of transmission coefficients are intrinsic for the extended $\bf{ k\cdot p}$ equation, based on higher order $\bf{ k\cdot p}$ terms in the low-energy limit. Such high-order approximations  lead to predicting small backscattering due to the effects of trigonal warping \cite{Ando2} of the bands, destroying perfect transmission in the channel.

The paper is organized as follows. In the framework of the
nearest-neighbor tight-binding formalism we calculate energy
dispersion (Sec.\ref{2}) and eigenstates (Sec. \ref{3}) of electrons in the
honeycomb graphene lattice. The lattice, in contrast to Ref. \cite{Zheng}, is considered as a set of
rectangular elementary cells of four nonequivalent atoms (Fig.
\ref{yInfinite}). This assumption simplifies the description of electron
transport processes in GRs in the presence of step-like or
barrier-like site-energy profiles. Following Ref. \cite{Akhmerov}, in Section \ref{4} we derive boundary conditions
for infinite GRs with armchair edges. By
using a combination of early found eigenstates of
the honeycomb lattice, we obtain normalized wave solutions for
electrons in armchair GRs and formulate the quantization law for the
transverse component $k^m$ of the wave vector. As a result, the full
$\pi$-electron band spectra of graphene ribbon breaks into a set
of subbands, whose features are discussed in Section \ref{5}.

In Section \ref{6}, we find the group velocity and propagation direction of
an electron in an armchair ribbon. It is shown that the group
velocity is proportional to $\sin\theta$, where $\theta$ is
the electron phase shift between two neighboring unit cells of the
ribbon. In sections \ref{7} and \ref{8}, the transmission probabilities through
step-like and barrier-like potentials are calculated.
Based on the exact solutions, in Section \ref{9}, we
demonstrate that the transmission coefficients exhibit the new kind of resonance occurring when $\cos\theta=\cos\bar{\theta}$, where $\bar{\theta}$
is the inter-cell electron phase shift in the region where
the potential exists. The number and positions of the
transmission resonances are shown to be different in the regions
$k^m<2\pi/3$ and $k^m>2\pi/3$. As for one-mode propagation in semimetal GRs ($k^m=2\pi/3$), our theory predicts no unit propagation. In Section \ref{10}, by expanding
the expressions for $\cos\theta$, $\cos\bar{\theta}$ in the vicinity
of $E=0$, we explain why the Klein paradox does not hold in armchair GRs.

In Sec. \ref{11}, we discuss through-step and
through-barrier probabilities and provide some examples. Conclusions are
presented in Sec. \ref{12}. Appendices \ref{A} (on electron
flux in the tight-binding model) and \ref{B} (on through-step transmission
in a linear chain of identical atoms) describe some theoretical tools used in the paper. Appendix \ref{C} contains some technical details related to obtaining our approximate solution from the Dirac solution \cite{Kats}.

\section{Dispersion relation for honeycomb graphene lattice}\label{2}

\begin{figure}
\centering
\includegraphics[width=0.38\textwidth]{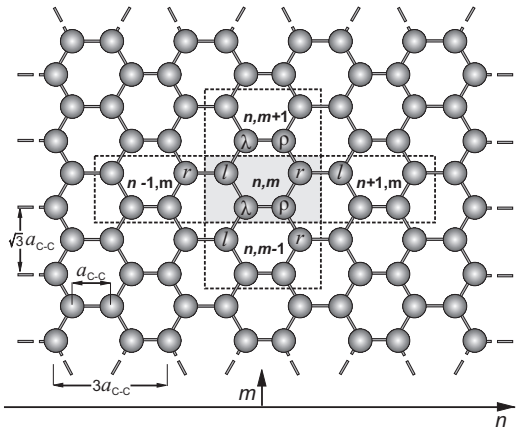}
\caption{Sketch of 2D honeycomb lattice. The central shadowed
block corresponds to an elementary cell $\{n,m\}$ consisting of 4
carbon atoms, labeled as $l,\lambda,\rho,r$. The lengths $3a_{\rm
C-C}$ and $\sqrt{3}a_{\rm C-C}$ are the translation periods in $n$
and $m$ directions, $a_{\rm C-C}$ is ${\rm C-C}$ bond
length.}\label{yInfinite}
\end{figure}

We consider the graphene honeycomb lattice as a set of
rectangular elementary cells of four atoms
$\alpha=l,\lambda,\rho,r$ (see Figure \ref{yInfinite}). By taking the
tight-binding representation for molecular orbitals
$$|\Psi\rangle=\sum_{n,m=-\infty}^{\infty}\sum_{\alpha}\psi_{n,m,\alpha}|n,m,\alpha\rangle,
$$ we come to the set of linear Schr\"{o}dinger equations
\begin{equation}\label{A1}
\left\{\!\!
\begin{array}{l}
-E\psi_{n,m,l}=\psi_{n,m,\lambda}+\psi_{n,m+1,\lambda}+\psi_{n-1,m,r},\\
-E\psi_{n,m,\lambda}=\psi_{n,m,l}+\psi_{n,m,\rho}+\psi_{n,m-1,l},
\\
-E\psi_{n,m,\rho}=\psi_{n,m,\lambda}+\psi_{n,m,r}+\psi_{n,m-1,r},
\\
-E\psi_{n,m,r}=\psi_{n,m,\rho}+\psi_{n,m+1,\rho}+\psi_{n+1,m,l},
\end{array}
\right.
\end{equation}
with respect to wave function components
$\psi_{n,m,\alpha}=\langle \Psi|n,m,\alpha\rangle$. Here
$|n,m,\alpha\rangle$ is $2{\rm p}_z$ orbital of $\alpha$-th atom
in $\{n,m\}$-elementary cell, $E\equiv E/|\beta|$ is the electron
energy in units of $|\beta|$, $\beta<0$ is a transfer integral
between the nearest-neighbor carbon atoms. Site-energies of
carbons equal zero and serve as a reference.

We look for a solution of Eq. (\ref{A1}) in the form
\begin{equation}\label{A2}
\psi_{n,m,\alpha}=\phi_{\alpha}e^{ik^n n+ik^m m}, \quad \alpha=
l,\lambda,\rho,r,
\end{equation}
where the dimensionless wave numbers $k^n(\equiv 3k_{x}a_{\rm
C-C})$, $k^m(\equiv \sqrt{3}k_{y}a_{\rm C-C})$, and
co-factors $\phi_{\alpha}$ are unknown coefficients,
$k_{x}$ and $k_{y}$ are the components of the wave vector along $n$ and $m$
directions respectively. The lengths $3a_{\rm C-C}$ and
$\sqrt{3}a_{\rm C-C}$ are the translation periods of
the graphene sheet as depicted on the Fig. \ref{yInfinite}, where
$a_{\rm C-C}$ is the C-C bond length.

Plugging in the solutions (\ref{A2}) into the system (\ref{A1}) gives the system of linear
equations
\begin{equation}\label{A3}
\left\{\!\!
\begin{array}{l}
E\phi_{l}+\left(1+e^{ik^m}\right)\phi_{\lambda}+
e^{-ik^n}\phi_{r}=0,\\ \left(1+e^{-ik^m}\right)\phi_{l}+
E\phi_{\lambda}+\phi_{\rho}=0,\\
\phi_{\lambda}+E\phi_{\rho}+\left(1+e^{-ik^m}\right)\phi_{r}=0,\\
e^{ik^n}\phi_{l}+\left(1+e^{ik^m}\right)\phi_{\rho}+E\phi_{r} =0.
\end{array}
\right.
\end{equation}
A nontrivial solution of Eq. (\ref{A3}) demands a zero determinant.
It leads to the dispersion relation
\begin{equation}\label{A4}
\cos{k^n}=f(E,k^m),\qquad f(E,k^m)=\frac{\left[E^2-1-4\cos^2
\frac{k^m}{2}\right]^2}{8\cos^2\frac{k^m}{2}}-1,
\end{equation}
defining the non-dimensional wave number $k^n$ in terms of $k^m$
and electron energy $E$. It follows from Eq.(\ref{A4}) that
\begin{equation}\label{A5}
E^2=1\pm 4\cos\frac{k^m}{2}\cos\frac{k^n}{2}+4\cos^2
\frac{k^m}{2},
\end{equation}
where index $\pm$ distinguishes two branches of the dispersion
relation.

\section{Wave solution in graphene lattice}\label{3}

If the relation (\ref{A4}) holds, system (\ref{A3}) becomes
degenerated, and we can express the coefficients $\phi_{l}$,
$\phi_{\lambda}$, $\phi_{\rho}$ in terms of $\phi_{r}$, omitting the
fourth equation in (\ref{A3}). The result can be written as follows
\begin{equation}\label{A6}
\phi_{\alpha}=C\,\widetilde{\phi}_{\alpha},\qquad
\alpha=l,\lambda,\rho,r,
\end{equation}
where
$$
\widetilde{\phi}_{l}=-
\frac{\left(E^2-1\right)e^{-ik^n}+4\cos^2\frac{k^m}{2}}
{E\left(E^2-1-4\cos^2\frac{k^m}{2}\right)},\qquad \widetilde{\phi}_{\lambda}=\frac{\left(1+e^{-ik^m}\right)\left(1+e^{-ik^n}\right)}
{E^2-1-4\cos^2\frac{k^m}{2}},
$$
$$
\widetilde{\phi}_{\rho}=-
\frac{\left(1+e^{-ik^m}\right)\left(E^2+e^{-ik^n}-4\cos^2\frac{k^m}{2}\right)}
{E\left(E^2-1-4\cos^2\frac{k^m}{2}\right)},\qquad \widetilde{\phi}_{r}=1.
$$

Exploiting the energy dispersion relation (\ref{A4}), one can see that
$|\widetilde{\phi}_{\alpha}|=1$ for the real-valued $k^n$.
Introducing a new function
\begin{equation}\label{A10}
e^{i\theta}=-\frac{E^2-1+4e^{
ik^n}\cos^2\frac{k^m}{2}}{E\left(E^2-1-4\cos^2\frac{k^m}{2}\right)}=-\frac{1\pm
2\cos\frac{k^m}{2}\,e^{ik^n/2}}{E},
\end{equation}
we can get
\begin{equation}\label{A11}
\widetilde{\phi}_{l}=e^{i(\theta -k^n)},\quad
\widetilde{\phi}_{\lambda}=\pm e^{-i(k^m+k^n)/2},\quad
\widetilde{\phi}_{\rho}=\mp e^{-i(k^m+k^n)/2+i\theta },\quad
\widetilde{\phi}_{r}=1.
\end{equation}
The choice of upper/lower signs in (\ref{A10}), (\ref{A11}) is
determined by the sign in (\ref{A5}). Plugging in Eq.
(\ref{A11}) into Eq. (\ref{A6}), and then the resulting expression into
(\ref{A2}) gives the expressions for electron eigenstates in the honeycomb lattice
\begin{equation}\label{A12}
\psi_{n,m,\alpha}(k^n,k^m)=C\left\{\!\!
\begin{array}{ll}
e^{ik^n(n-1)+ik^m m+i\theta},&\alpha=l,\\ \pm
e^{ik^n(n-1/2)+ik^m(m-1/2)},&\alpha=\lambda,
\\
\mp e^{ik^n(n-1/2)+ik^m(m-1/2)+i\theta},&\alpha=\rho,\\ e^{ik^n
n+ik^m m},&\alpha=r.
\end{array}
\right.
\end{equation}

\section{Wave solution for armchair GRs}\label{4}

\begin{figure}
\centering
\includegraphics[width=0.42\textwidth]{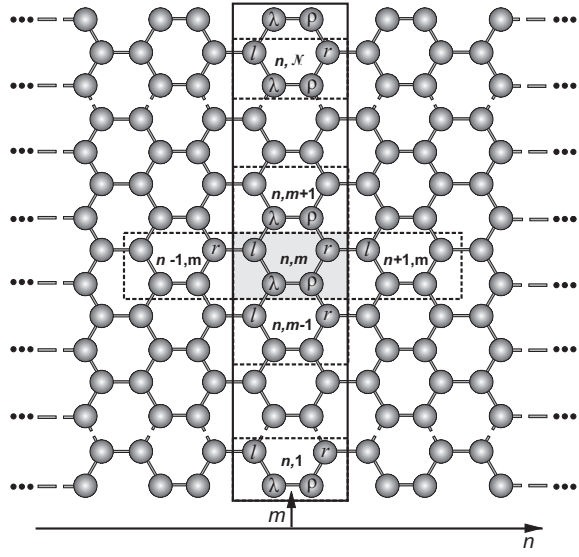}
\caption{Sketch of infinite graphene ribbon with  armchair edges.
The solid-framed block is the unit cell $n$ of GR.}\label{NRarmchair}
\end{figure}

Unlike the 2D honeycomb lattice, the electron wave function components
$\psi_{n,m,\alpha}$ in GRs with armchair edges (see Figure
\ref{NRarmchair}) are the solution to the linear Schr\"{o}dinger
equations (\ref{A1}) only for inner elementary cells ($1<m\leq {\cal N}$). For boundary cells with $m=1$ or $m={\cal
N}+1$, one needs to take into account that boundary carbons have only two
neighbors (see Fig. \ref{NRarmchair}).
It demands the wave function to vanish on the set of absent sites \cite{Malysheva,Akhmerov}$^,$\footnote{A more detailed description may be found here: Yu. Klymenko and O. Shevtsov, arXiv:0806.4531v1 [cond-mat.mes-hall].}
\begin{equation}\label{A17}
\psi_{n,0,l}=0, \quad \psi_{n,0,r}=0,\quad \psi_{n,{\cal
N}+1,l}=0,\quad \psi_{n,{\cal N}+1,r}=0.
\end{equation}
Similar boundary conditions were used in Refs \cite{Robinson, Zheng}.

To satisfy the relations (\ref{A17}), we represent the solution as a  linear combination of the states (\ref{A12}),
\[
\psi_{n,m,\alpha}(k^n)=\psi_{n,m,\alpha}(k^n,k^m)-\psi_{n,m,\alpha}(k^n,-k^m).
\]
Since the definition
(\ref{A10}) implies that $\theta(-k^m)=\theta(k^m)$, we finally obtain
\begin{equation}\label{A19}
\psi_{n,m,\alpha}(k^n)=C'\left\{\!\!
\begin{array}{ll}
e^{ik^n(n-1)+i\theta}\,\sin k^m m,&\alpha=l, \\ \pm
e^{ik^n(n-1/2)}\,\sin k^m (m-1/2),&\alpha=\lambda,\\\mp
e^{ik^n(n-1/2)+i\theta}\,\sin k^m (m-1/2),&\alpha=\rho,\\
e^{ik^n n}\,\sin k^m m,&\alpha=r,
\end{array}
\right.
\end{equation}
where
\begin{equation}\label{A20}
k^m\equiv k^m_j=\frac{\pi j}{{\cal N}+1},\qquad j=1,2,\ldots, {\cal N},
\end{equation}
is the set of discretized transversal wave numbers, and the unknown
constant $C'(\equiv 2iC)$ can be found from the normalization
condition over a unit cell of the armchair GR,
\[
\sum_{m=1}^{{\cal
N}}\left\{\left|\psi_{n,m,l}\right|^2+\left|\psi_{n,m,r}\right|^2\right\}+\sum_{m=1}^{{\cal
N}+1}\left\{\left|\psi_{n,m,\lambda}\right|^2+\left|\psi_{n,m,\rho}\right|^2\right\}=1.
\]
Then we finally get
$C'=[2({\cal N}+1)]^{-1/2}$.

\section{Band spectra of armchair GRs}\label{5}

Due to the transverse momentum quantization (\ref{A20}), the full
$\pi$-electron band spectra of an armchair GR breaks into a
set of subbands connected with each mode $j$ independently. Given $E$ and $k^m=k^m_j$, the dispersion relation (\ref{A4})
allows us to establish some important peculiarities of the $j$-th part
of the band spectra. Following the theory of
a one-dimensional crystal \cite{Onipko} with an arbitrary electronic structure
of the elementary cell, each $j$-th band of graphene
electron spectra includes 4 subbands, symmetrically disposed with
respect to $E=0$. Real values of longitudinal wave number $k^n\in
[0,\pi]$ define the region of propagating electron states (when $|f(E,k^m_j)|<1$ in Eq.(\ref{A4})). The cases $f(E,k^m_j)>1$ and
$f(E,k^m_j)<-1$ are related to the complex values of $k^m=i\delta$ and
$k^m=\pi+i\delta$, respectively. They refer to the forbidden zones
-- the gaps between neighboring electron subbands and the regions
above the highest subband and below the lowest one. The subband
boundaries correspond to $k^m=0$ or $ k^m=\pi$ and satisfy the solutions to
$f(E,k^m_j)=1$ or $f(E,k^m_j)=-1$, respectively. Since Eq.(\ref{A4}) defines only one $k^m$ for each energy level $E$, the subbands of the $j$th band cannot overlap, however,
they may touch each other along the frontiers with the same
$k^m(=0,\pi)$.

\begin{figure}[htb]
\includegraphics[width=\textwidth]{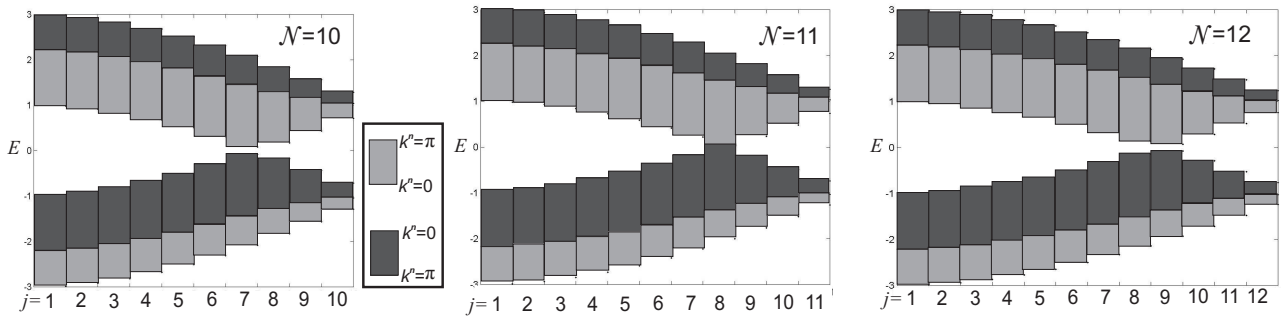}
\caption{Examples of energy bands for armchair graphene ribbons
with ${\cal N}=10-12$. The transverse momentum quantization
results in subband series connected with each mode $j$
independently. Gray (black) rectangles mark subbands, whose bottom
corresponds to $k^n=0$ ($k^n=\pi$) and whose top -- to $k^n=\pi$
($k^n=0$), as depicted in the framed panel. The border between
black and gray rectangles in the positive- and negative-energy
spectra  determines the subband connecting $k^n=\pi$ frontiers,
this is a degenerated solution to $f(E,k^m_j)=-1$. The
fundamental gap of a graphene ribbon is the minimal gap value
$\Delta$ between the valence and conduction
bands among all modes $j=1,\ldots,{\cal N}$.
}\label{GRsArm10-12}
\end{figure}

The results of the band spectrum modeling for armchair GRs with ${\cal
N}=10-12$ are represented in Figure \ref{GRsArm10-12}. Similar shapes
are observed for GRs with arbitrary number ${\cal N}$.
The common features of the band spectra are: (I) bands are bounded
by the energy interval $|E|\leq 3$; (II) the band structure is
symmetric with respect to $E=0$; (III) if the number ${\cal N}+1$
is divisible by 3, the mode $j=2({\cal N}+1)/3$ (or
$k^m_j=2\pi/3$) does not possess the energetic gap between
positive- and negative-energy bands, and, therefore, it possesses the semimetal type
of electron conductivity \cite{Brey,Malysheva,Robinson,Zheng}. For other
$j$-modes there is a gap
\begin{equation}\label{A22a}
\Delta=2\left|1-2\cos\frac{k^m}{2}\right|
\end{equation}
between positive- and negative-energy subbands, which corresponds to
a semiconducting ribbon whose gap decreases with increasing $\cal
N$.

\section{Propagation velocity}\label{6}

To obtain propagation velocity of the wave, we plug in  the solution (\ref{A19})
into the relation (A2) for the electron flux $J$. Then
$J=|\beta|\sin\theta/2\hbar$.
Using the definition (\ref{A10}) and
the first derivative of the dispersion relation
(\ref{A4}) with respect to $E$, one can get
\[
J=-\frac{|\beta|}{\hbar}\,\frac{2\sin k^n\cos^2
\frac{k^m}{2}}{E\left(E^2-1-4\cos^2
\frac{k^m}{2}\right)}=\frac{|\beta|}{\hbar}\,\frac{dE}{dk^n}.
\]
Representing the flux $J$ as $(|\beta|/\hbar)v(k^n)$, where
$v(k^n)$ is the dimensionless group velocity, we finally obtain
\begin{equation}\label{A25}
v(k^n)=\frac{dE}{dk^n}=\frac{\sin\theta}{2}.
\end{equation}

The sign of $v$ in Eq. (\ref{A25}) determines the moving direction of the propagating wave
(\ref{A19}). For a subband $\left(=_0^{\pi}\right)$, whose bottom
corresponds to $k^n=0$ and top to $k^n=\pi$ (see
Fig. \ref{GRsArm10-12}), the wave number $k^n$ increases along with
$E$ ($dE/dk^n>0$). Therefore, $v>0$ and the state (\ref{A19}) describes the right-moving mode.

Similarly, in a subband of type $\left(=^0_{\pi}\right)$, we
have $dE/dk^n<0$, and the wave solution (\ref{A19}) propagates to the left.
Replacing $k^n$ with $-k^n$ in (\ref{A10}), (\ref{A19}), we
obtain the expression describing a right-moving mode in these subbands
\begin{equation}\label{A26}
\psi_{n,m,\alpha}(-k^n)=C'\left\{\!\!
\begin{array}{ll}
e^{-ik^n(n-1)-i\theta}\,\sin k^m m,&\alpha=l, \\ \pm
e^{-ik^n(n-1/2)}\,\sin k^m (m-1/2),&\alpha=\lambda,\\\mp
e^{-ik^n(n-1/2)-i\theta}\,\sin k^m (m-1/2),&\alpha=\rho,\\
e^{-ik^n n}\,\sin k^m m,&\alpha=r.
\end{array}
\right.
\end{equation}

From (\ref{A19}) and (\ref{A26}), one can understand the physical
meaning of the quantity $\theta$. Expressing $\psi_{n+1,m,l}(\pm
k^n)$ via $\psi_{n,m,r}(\pm k^n)$, we get
$\psi_{n+1,m,l}(\pm k^n)=e^{\pm i\theta}\psi_{n,m,r}(\pm
k^n)$, i.e., an electron, propagating through the corresponding
subbands, gains phase shift $\pm\theta$ between two neighboring
unit cells of GR. The phase shift $\theta$ plays an important role
in electron transmission through the ribbons, as discussed below.
Also note that $\sin\theta>0$ inside the $\left(=_0^{\pi}\right)$-subbands and $\sin\theta<0$ in the subbands
$\left(=^0_{\pi}\right)$, which is also seen from the relation
(\ref{A25}).

The derived analytical form of the wavefunctions (\ref{A19}), (\ref{A26}) and the expression (\ref{A25}) for propagation velocity will be used below to investigate transport properties of armchair GRs in the presence of step-like or barrier-like potentials.

\section{Transmission probability through a potential step}\label{7}

In this section, we derive the transmission coefficient for an electron with
energy $E$, transverse and longitudinal wave numbers $k^m$, $k^n$,
meeting the electrostatic potential $$V(n)=\left\{\!\!
\begin{array}{ll}
0,&n\leq 0, \\ V_{0},& n\geq 1,
\end{array}
\right. $$ in an armchair ribbon. We denote the wavefunction by $\psi^{^{\rm
left}}_{n,m,\alpha}$ for $n\leq 0$ and $\psi^{^{\rm
right}}_{n,m,\alpha}$ for $n \geq 1$. Let $U_0\equiv
eV_0/|\beta|$.
\begin{figure}
\centering
\includegraphics[width=0.4\textwidth]{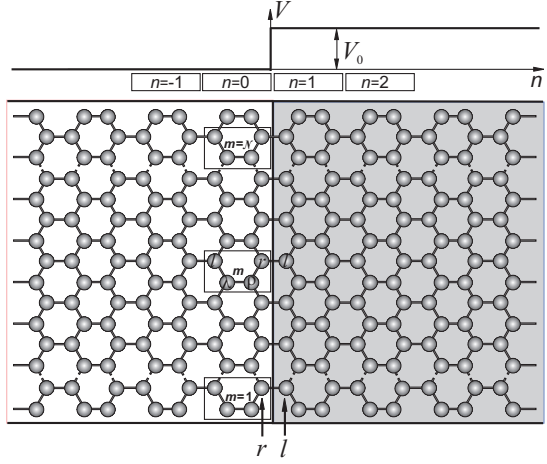}
\caption{Sketch of a graphene ribbon with  armchair edges modified
by a step-like potential. Unit cells with $n\leq 0$ and $n \geq 1$
belong to the left and right leads of GR, respectively.}
\label{StepArmGRs}
\end{figure}

The Schr\"{o}dinger equations for all atoms in both leads are used to obtain the probabilities of electron scattering through the
potential step. The comparison of these equations near the interface between two regions ($n=0,1$) allows us to write down the matching conditions$^{26}$ as follows
\begin{equation}\label{A29}
\psi^{^{\rm left}}_{0,m,r}=\psi^{^{\rm right}}_{0,m,r},\quad
\psi^{^{\rm left}}_{1,m,l}=\psi^{^{\rm right}}_{1,m,l},\qquad
m=1,\ldots,{\cal N}.
\end{equation}
Similar conditions for the tight-binding model were used in Ref. \cite{Shomerus} and for the Dirac equation in Refs. \cite{Tworzydlo, Kats, Kats2}.

\begin{figure}
\centering
\includegraphics[width=0.3\textwidth]{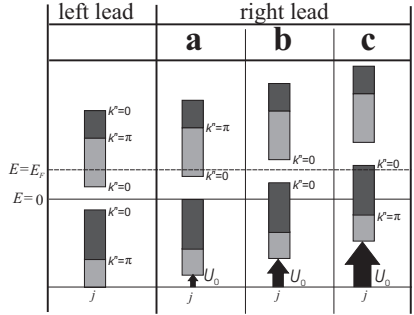}
\caption{Three possible types of electron subband disposition in
the right region. The Fermi level is assumed to be located in a
$\left(=_0^{\pi} \right)$-subband of the left lead and incident
wave is described by the solution (\ref{A19}). {\bf a)} Electron
propagates through a $\left(=_0^{\pi} \right)$-subband in the
right region; {\bf b)} the Fermi level belongs to the band-gap of
the right lead, which results in total reflection; {\bf c)}
electron propagates through a $\left(=^0_{\pi} \right)$-subband in
the right region. This case corresponds to the solution (\ref{A26})
for an outgoing wave in the lead.}
\label{StepGRs3pos}
\end{figure}

Since the matching conditions (\ref{A29}) do not mix the modes, we need to sew
only the $j$-th solutions in each region under study. Wave solutions,
describing incident and reflected electron waves in the left lead,
can be constructed from the states (\ref{A19}) and (\ref{A26}).
For the right region, one needs to take into account the shift of site energies produced by the applied potential. Introducing additional
notation for longitudinal wave number $\bar{k}^n$, phase
$\bar{\theta}$ and group velocity $\bar{v}(\bar{k}^n)$ in the
right lead,
\begin{equation}\label{A30a}
\bar{k}^n\equiv k^n(E\rightarrow E-U_{0}),\quad
\bar{\theta}\equiv\theta(E\rightarrow E-U_{0},k^n\rightarrow
\bar{k}^n),\quad \bar{v}(\bar{k}^n)\equiv
\frac{\sin\bar{\theta}}{2},
\end{equation}
we first consider the case of propagation through the
$\left(=_0^{\pi} \right)$-subbands in both regions, as depicted on
the {\bf a}-panel of Fig. \ref{StepGRs3pos}.
The solutions describing the incident and reflected electron waves in
the left lead and the outgoing wave in the right region
should be taken in the following form
\begin{equation}\label{A30}
\psi^{^{\rm left}}_{n,m,l}=\left[e^{ik^n(n-1)+i\theta}+
re^{-ik^n(n-1)-i\theta}\right]\sin k^m_j m, \quad \psi^{^{\rm
left}}_{n,m,r}=\left[e^{ik^n n}+re^{-ik^n n}\right]\sin k^m_j m,
\end{equation}
\begin{equation}\label{A31}
\psi^{^{\rm right}}_{n,m,l}=te^{i\bar{k}^n(n-1)+i\bar{\theta}}\sin
k^m_j m, \quad \psi^{^{\rm right}}_{n,m,r}= te^{i\bar{k}^n n}\sin
k^m_j m,
\end{equation}
where $r$ and $t$ are the amplitudes of the reflected and transmitted
waves excited by the incident wave with transverse mode $j$. (We omit
the factor $C'$ for simplicity.) After plugging in \ref{A30}),
(\ref{A31}) into the conditions (\ref{A29}), we get the following
two equations $$
\left\{\!\!
\begin{array}{l}
1+r=t,\\ e^{i\theta}+re^{-i\theta}=te^{i\bar{\theta}}.
\end{array}
\right. $$ Their solution is
\begin{equation}\label{A32}
r=-\frac{e^{i\bar{\theta}}-e^{i\theta}}{e^{i\bar{\theta}}-e^{-i\theta}},\qquad
t=\frac{2i\sin\theta}{e^{i\bar{\theta}}-e^{-i\theta}}.
\end{equation}

The reflection and transmission probabilities $R$ and $T$ can be directly found from the
scattering amplitudes $r,t$, using their definitions: $R=|r|^2$
and
$T=|t|^2\bar{v}(\bar{k}^n)/v(k^n)=|t|^2\sin\bar{\theta}/\sin\theta$.
In this way, we obtain the expression for an electron propagating via $\left(=_0^{\pi} \right)$-subbands in both leads
\[
T=\frac{4\sin\theta\,\sin\bar{\theta}}
{4\sin\theta\,\sin\bar{\theta}+\left(\sin\bar{\theta}-
\sin\theta\right)^2+ \left(\cos\bar{\theta}-\cos\theta\right)^2}.
\]

The {\bf b}-panel of Figure \ref{StepGRs3pos} corresponds to
the case when the electron energy belongs to the forbidden band in
the right region (it is impossible for one-mode propagation through a semimetal ribbon with a zero-gap,
$k^m_j=2\pi/3$). In this case, the wave number $\bar{k}^n$ takes imaginary values, and the
expression for $t$ in (\ref{A32}) is to be omitted. At the same
time, for imaginary-valued $\bar{k}^n$ the expression for
$e^{i\bar{\theta}}$ becomes real, which results in $R=1$ after using
the $r$-solution from (\ref{A32}).

For an electron moving through a $\left(=^0_{\pi} \right)$-subband in
the right lead, the {\bf c}-panel of Figure \ref{StepGRs3pos}, we
use the solution (\ref{A26}) for an outgoing wave (see the
definitions (\ref{A30a})). Thus, we obtain
\[
T=\frac{-4\sin\theta\,\sin\bar{\theta}}
{-4\sin\theta\,\sin\bar{\theta}+\left(\sin\bar{\theta}+
\sin\theta\right)^2+ \left(\cos\bar{\theta}-\cos\theta\right)^2},
\]
where $\sin\bar{\theta}<0$, as stated in Sec.\ref{6}. In a
similar way, one can determine the transmission coefficient for
an electron propagating through arbitrary subbands in the left and
right leads
\begin{equation}\label{A35}
T^{\rm step}=\frac{4|\sin\theta\,\sin\bar{\theta}|}
{4|\sin\theta\,\sin\bar{\theta}|+\left(|\sin\bar{\theta}|-
|\sin\theta|\right)^2+
\left(\cos\bar{\theta}-\cos\theta\right)^2}.
\end{equation}

The common relation (\ref{A35}) must be supplemented with the expressions for $\cos\theta$, $\cos\bar{\theta}$ determined from
(\ref{A10}) and (\ref{A30a}). Then, from the energy
dispersion (\ref{A4}), we finally obtain
\begin{equation}\label{A36}
\cos\theta=g(E,k^m),\qquad \cos\bar{\theta}=g(E-U_{0},k^m),
\end{equation}
where the function $g$ is described by the simple relation
\begin{equation}\label{A37}
g(E,k^m)=-\frac{E^2+1-4\cos^2\frac{k^m}{2}}{2E}.
\end{equation}
The analytical expression (\ref{A35}) with the definitions
(\ref{A36}), (\ref{A37}) is one of the central points of this
paper since it makes it possible to compute the
transmission coefficient through a potential step without recourse
to the initial dispersion relation (\ref{A4}).

It is worth noting how similar the
solution (\ref{A35}) to the relation (B4) is, though the relation (B4) is for through-step transmission coefficient in the 1D undimerized chain of identical atoms. In other words, the inter-cell variables $\theta$,
$\bar{\theta}$ for GRs act as wave numbers $k$, $\bar{k}$ in the
linear chain model, whose unit cell degenerates to a zero-sized
atom. We'll return to this issue in Sec. \ref{9}.

\section{Transmission coefficient through a potential barrier}\label{8}

\begin{figure}
\centering
\includegraphics[width=0.45\textwidth]{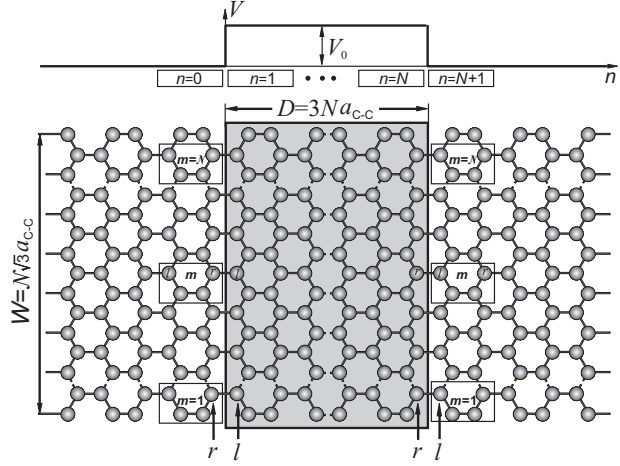}
\caption{Sketch of a graphene ribbon with  armchair edges
modified by the barrier-like potential. $W$ and $D$ are the ribbon
width and length of the disturbed region, respectively.}
\label{BarArmGRs}
\end{figure}

To obtain the exact expression for the electron transmission
probability through a potential barrier $$V(n)=\left\{\!\!
\begin{array}{ll}
0,&n\leq 0, \\ V_{0},&1\leq n\leq N,\\0,&n\geq N+1,
\end{array}
\right. $$ (see Fig. \ref{BarArmGRs}), we denote the wavefunction by $\psi^{^{\rm
left}}_{n,m,\alpha}$ for the left lead ($n\leq 0$), $\psi^{^{\rm
in}}_{n,m,\alpha}$ for the "in" region ($1\leq n \leq N$) perturbed
by the potential, $\psi^{^{\rm right}}_{n,m,\alpha}$ for
the right lead ($n \geq N+1$), and keep the same notation for the
disturbed region, as in the previous section.
By analogy with the matching conditions (\ref{A29}), we need to have the following relations satisfied
\begin{equation}\label{A38}
\begin{array}{ll}
\psi^{^{\rm left}}_{0,m,r}=\psi^{^{\rm in}}_{0,m,r},& \psi^{^{\rm
left}}_{1,m,l}=\psi^{^{\rm in}}_{1,m,l},\\ \psi^{^{\rm
in}}_{N,m,r}=\psi^{^{\rm right}}_{N,m,r},& \psi^{^{\rm
in}}_{N+1,m,l}=\psi^{^{\rm right}}_{N+1,m,l}.
\end{array}
\end{equation}
As before, the solutions are determined using the states (\ref{A19}) and (\ref{A26}).

For an electron propagating through a
$\left(=_0^{\pi} \right)$ subband in the left lead the wavefunction expressions are presented by (\ref{A30}). The outgoing wave in the right lead
is constructed from the state (\ref{A19}),
\[
\phi^{{\rm right}}_{n,m,\alpha}=t\sin (k^m_j m)\,\left\{\!\!
\begin{array}{ll}
e^{ik^n (n-1)+i\theta},&\alpha=l,
\\e^{ik^n n},&\alpha=r.
\end{array}
\right.
\]
For the in-region we determine a solution
\begin{eqnarray*}
\psi^{{\rm in}}_{n,m,\alpha}=\sin (k^m_j m)\left\{\!\!
\begin{array}{ll}
\left(ae^{i\bar{k}^n(n-1)+i\bar{\theta}}+be^{-i\bar{k}^n(n-1)-
i\bar{\theta}}\right),&\alpha=l,
\\\left(ae^{i\bar{k}^n
n}+be^{-i\bar{k}^n n}\right),&\alpha=r,
\end{array}
\right.
\end{eqnarray*}
where $a$ and $b$ are additional unknown constants. The solution does not depend on the subband type of propagation in the disturbed region. The case of tunneling through a gap, with $\bar{k}^n$ imaginary, is not considered in the paper.

Using the matching conditions (\ref{A38}), we obtain the following four
equations with four unknowns $r$, $t$, $a$ and $b$
\begin{equation}\label{A41}
\left\{
\begin{array}{l}
a+b=1+r,\\
e^{i\theta}+re^{-i\theta}=ae^{i\bar{\theta}}+be^{-i\bar{\theta}},\\
t\,e^{ik^n N}=ae^{i\bar{k}^nN}+be^{-i\bar{k}^nN},\\
ae^{i\bar{k}^nN+i\bar{\theta}}+
be^{-i\bar{k}^nN-i\bar{\theta}}=te^{ik^n N+i\theta}.
\end{array}
\right.
\end{equation}
Solving (\ref{A41}) with respect to $t$, we obtain
\[
t=-\frac{2i g_{\rm nd}\sin\theta }{\left(g_{\rm
d}-e^{-i\theta}\right)^2-g^{2}_{\rm nd}}\,e^{-ik^n N-i\theta},
\]
where the functions
\begin{equation}\label{A42}
g_{\rm d}=\frac{\sin
\bar{k}^nN}{\sin(\bar{k}^nN+\bar{\theta})},\qquad g_{\rm
nd}=\frac{\sin\bar{\theta}}{\sin(\bar{k}^nN+\bar{\theta})}
\end{equation}
depend only on the "disturbed" variables $\bar{k}^n$ and $\bar{\theta}$. Since, by definition,
$T=|t|^2$, the through-barrier transmission probability can be written in the following compact
form$^{26}$
\begin{equation}\label{A44}
T^{\rm bar}=\frac{\sin^2\theta\,\sin^2\bar{\theta}}{\sin^2\theta
\sin^2\bar{\theta}+\left(\cos\theta-\cos\bar{\theta}\right)^2\sin^2(\bar{k}^nN)}.
\end{equation}
It is easy to verify that
the transmission coefficient (\ref{A44}) remains unchanged for propagation
through a $\left(=^0_{\pi} \right)$-subband in external (left and
right) leads.

As follows from Eq. (\ref{A44}), the unit
transmission occurs under the coincidence of $\cos\theta$ and
$\cos\bar{\theta}$. This can be viewed as a new type of resonance, which differs from the familiar resonance condition $\sin(\bar{k}^nN)=0$. Some of its properties are studied in the next
section.

\section{Unit propagation condition}\label{9}

The transmission
coefficients (\ref{A35}) and (\ref{A44}) through step-like and barrier-like potentials
exhibit the unit propagation at $\theta$s such that $\cos\theta=\cos\bar{\theta}$, which are the roots
of the equation
\begin{equation}\label{A52}
g(E,k^m)=g(E-U_0,k^m),
\end{equation}
(see the definitions (\ref{A36})). Particularly, Eq.(\ref{A52}) has
no solution for a function $g(E,k^m)$ that is monotonic with respect to $E$, except when $U_0=0$, the usual condition of the unit transmission.

\begin{figure}
\centering
\includegraphics[width=0.4\textwidth]{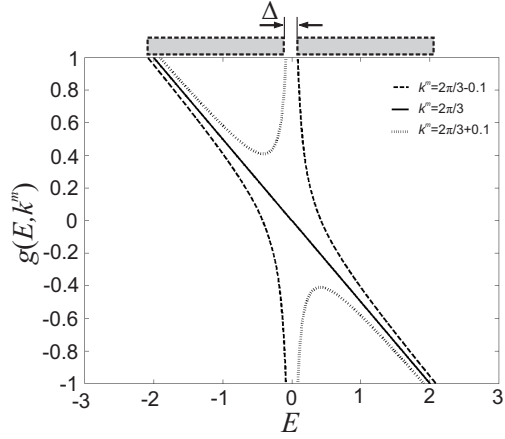}
\caption{Behaviour of $g(E,k^m)$ inside the positive-
and negative-energy bands for $k^m=2\pi/3$ (solid line) and
$k^m=2\pi/3\mp 0.1$. The shaded rectangles represent the band positions for
the transverse wave number $k^m=2\pi/3-0.1$, where $\Delta$ is the
energy gap for given $k^m$.} \label{Theta}
\end{figure}

For the case of one-mode propagation in semimetal GRs ($k^m=2\pi/3$) the relation
(\ref{A37}) reduces to the simple linear dependence
\begin{equation}\label{A54}
g(E,k^m)=-E/2,
\end{equation}
which corresponds to the absence of perfect transmission. In cases of one-mode propagation in semiconducting GRs
($k^m\ne 2\pi/3$), due to the oddness of $g$ with respect to $E$, we
confine our attention to the properties of $g(E,k^m)$ inside
the positive-energy region $E>0$. Differentiating $g(E,k^m)$ with respect to $E$,
\[
\frac{dg(E,k^m)}{dE}\equiv
-\frac{1}{2}+\frac{1-4\cos^2\frac{k^m}{2}}{2E^2}=0,\qquad
d^2g/dE^2>0,
\]
one can compute the point where the function attains its minimum value
\begin{equation}\label{A53a}
E=E_{min}(k^m)=\sqrt{1-4\cos^2\frac{k^m}{2}},
\end{equation}
only for transverse wave numbers in the region
$2\pi/3<k^m\leq \pi$. For the case when $0<k^m<2\pi/3$, we
obtain $dg/dE\ne 0$ and $d^2g/dE^2<0$. Hence
$g(E,k^m)$ is convex and monotone decreasing. Figure
\ref{Theta} presents the in-band behaviour of $g(E,k^m)$ with respect to $E$ for the
above mentioned $k^m$-domains: $k^m=2\pi/3-0.1(<2\pi/3)$,
$k^m=2\pi/3$, and $k^m=2\pi/3+0.1(>2\pi/3)$.

As follows from the relation (\ref{A52}), the transmission probabilities (\ref{A35}) and
(\ref{A44}) reach unity at the points where the curve $g(E,k^m)$
crosses $g(E-U_0,k^m)$. Such intersections are possible even for $U_0\ll 1$ in the region $2\pi/3<k^m\leq \pi$  (see Figure \ref{Theta}), where the resonance occurs near the
energy level (\ref{A53a}). In the case when $0<k^m<2\pi/3$, unit
propagation takes place if the value of $U_0$ exceeds the energy gap
$\Delta$.
To show this, we plug in the relation (\ref{A37}) into Eq. (\ref{A52}) and get
\begin{equation}\label{A55}
E(E-U_{0})=x,\qquad x=1-4\cos^2\frac{k^m}{2}.
\end{equation}
The unit transmission points are real-valued roots
of the Eq. (\ref{A55}),
\begin{equation}\label{A56}
E^{\rm res}=U_{0}/2\pm\sqrt{(U_0/2)^2+x}.
\end{equation}
In the interval $0<k^m<2\pi/3$, $x$ is negative, and
the transmission resonances exist only if $|U_0|$
exceeds $2|x|^{1/2}$. Expressing $x$ in terms of the energy gap
$\Delta$ (see Eq.(\ref{A22a})), we obtain the following inequality
for transmission resonances in this region
\begin{equation}\label{A56a}
|U_0|\geq 4\sqrt{\Delta(\Delta+1)}.
\end{equation}

\begin{figure}
\centering
\includegraphics[width=0.75\textwidth]{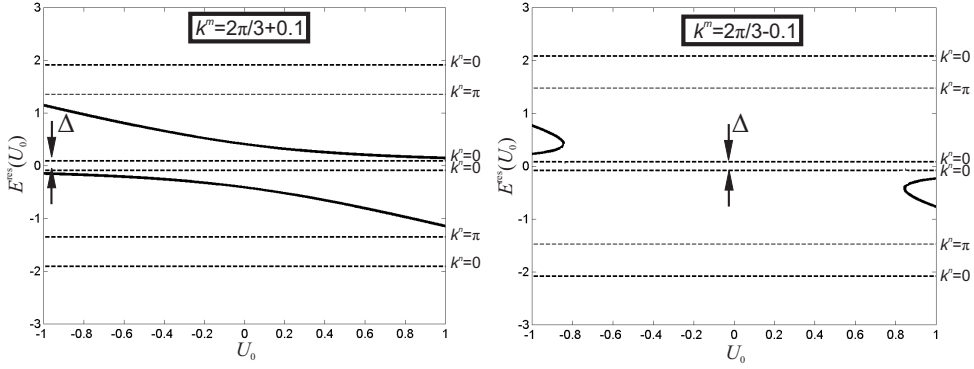}
\caption{Resonance energies $E^{\rm res}$ (solid lines) versus
$U_0$ for transverse wave numbers $k^m=2\pi/3\pm 0.1$. Dashed
horizontal lines correspond to the boundaries of the positive- and
negative energy subbands for the left region, where $\Delta$ is the
energy gap for given $k^m$.} \label{ResPosPRB}
\end{figure}

The energies of the transmission resonances versus $U_0$ are
plotted in Figure \ref{ResPosPRB}. If $2\pi/3<k^m\leq \pi$, the
indicated region $|U_0|\leq 1$ contains two resonances, one of
which belongs to the valence band, and another to the
conduction band. In the case when $0<k^m<2\pi/3$, two unit transmissions
are observed if the condition (\ref{A56a}) holds.
The resonance positions depend on the sign of $U_0$. Both of them belong to the valence band if $U_0>0$ and to the conduction band if
$U_0<0$.

The case of one-mode propagation in semimetal ribbons ($k^m=2\pi/3$) deserves additional
attention. The linear dependence of $g(E,k^m)$ with respect to $E$ (given by Eq. (\ref{A54})), makes unit
propagation impossible. This shows that the
Klein paradox \cite{Kats} does not hold in armchair GRs. Specifically, plugging in (\ref{A36}) and (\ref{A54}) into (\ref{A35}) produces
\begin{equation}\label{A57a}
T=\frac{4\sqrt{(4-E^2)[4-(E-U_0)^2]}}
{\left(\sqrt{4-E^2}+\sqrt{4-(E-U_0)^2}\right)^2+U_0^2},
\end{equation}
which can be reduced to
\begin{equation}\label{A57}
T\approx 1-U_0^2/16
\end{equation}
for $E=0$ and $|U_0|\ll 1$.

It is worth noting that the through-step
expression (\ref{A57a}) coincides with its analog for
electron transmission through a potential step in an undimerized linear
chain of identical atoms (see Appendix B). Specifically, the through-step coefficient (\ref{A57a}) in the vicinity of the neutrality point $E=0$ is exactly the same as the
transmission coefficient through a step potential formed in the linear chain when the electron energy is located near the corresponding band center.

\section{Low-energy limit}\label{10}

Most theoretical studies on the electronic properties of all-carbon
honeycomb lattices are based on the $\bf{ k\cdot p}$ approximation
in the vicinity of the zero-energy point, leading to the two-dimensional analog of the relativistic Dirac equation \cite{Geim,Ando,Brey,Kats}. Rewriting the energy dispersion (\ref{A5})
\[
E^2=\left(1\pm 2 \cos\frac{k^m}{2}\right)^2\mp
4\cos\frac{k^m}{2}\left(1-\cos\frac{k^n}{2}\right),
\]
we see that in the low-energy limit only the branch
\begin{equation}\label{A46}
E^2=\left(1- 2\cos\frac{k^m}{2}\right)^2+
4\cos\frac{k^m}{2}\left(1-\cos\frac{k^n}{2}\right)
\end{equation}
of the dispersion relation takes
part in electron transport. In terms of the deviation of ${\mathbf k}=(k^n,k^m)$ from
the zero-energy point $(0,2\pi/3)$,
\[
k^n=3k_{x}a_{\rm C-C},\qquad k^m=2\pi/3+\sqrt{3}k_{y}a_{\rm C-C},
\]
we can expand (\ref{A46}) into series $|k_{x(y)}|a_{\rm C-C }\ll 1$ up
to and including the quadratic terms. Then
\begin{equation}\label{A48}
E(k_x,k_y)=\pm\frac{3 a_{\rm C-C}}{2}\sqrt{k_x^2+k_y^2},
\end{equation}
which is similar to the cone-like form of electron dispersion
\cite{Geim,Ando,Malysheva}. In the same way, the
formula (\ref{A10}) for inter-cell electron phase shift $\theta$ reduces
to
\begin{equation}\label{A49}
e^{i\theta}=-\frac{1-2\cos\frac{k^m}{2}\,e^{ik^n/2}}{E}= i\,{\rm
sgn}(E)\,\frac{k_x+ik_y}{\sqrt{k_x^2+k_y^2}}+O(k_x,k_y)+\ldots\,,
\end{equation}
where
\begin{equation}\label{A50}
O(k_x,k_y)=-\frac{a_{\rm C-C}}{4}\,{\rm
sgn}(E)\,\frac{3k_x^2+6ik_xk_y+k_y^2}{\sqrt{k_x^2+k_y^2}}.
\end{equation}
Retaining only the first term in Eq. (\ref{A49}),
we come to
\begin{equation}\label{A51}
e^{i\theta}=i\,{\rm sgn}(E)\,e^{i\phi},\qquad
\phi=\tan^{-1}\frac{k_y}{k_x},
\end{equation}
where $\phi$ is the known phase factor from low-energy theory
\cite{Kats,Kats2}. Expressing the transmission coefficients  (\ref{A35}) and (\ref{A44}) in terms of
$\phi$, $\bar{\phi}\equiv \phi(k_x\rightarrow \bar{k}_x)$, we
obtain the following approximate expressions
\begin{equation}\label{A58}
T^{\rm step}_{\rm appr}=\frac{4|\cos\phi\,\cos\bar{\phi}|}
{\left(|\cos\bar{\phi}|+ |\cos\phi|\right)^2+ \left[{\rm
sgn}(E-U_0)\,\sin\bar{\phi}-{\rm sgn}(E)\,\sin\phi\right]^2},
\end{equation}
\begin{equation}\label{A59}
T^{\rm bar}_{\rm
appr}=\frac{\cos^2\phi\cos^2\bar{\phi}}{\cos^2\phi\cos^2\bar{\phi}+
\left[{\rm sgn}(E-U_{0 })\sin\bar{\phi}-{\rm
sgn}(E)\sin\phi\right]^2\sin^2(\bar{k}_xD)},
\end{equation}
for step-like and barrier-like potentials in the low-energy limit. The length $D=3a_{\rm C-C}N$ in
(\ref{A59}) is the width of the barrier, as depicted in
Figure \ref{BarArmGRs}.

The through-barrier coefficient
(\ref{A59}) coincidents with the expression \cite{Kats} obtained in the framework of the Dirac approach (see details in Appendix \ref{C}). This coincidence is not surprising. However, no "Dirac" analog of the through-step coefficient (\ref{A58}) has been obtained so far.

Eqs.(\ref{A58}) and (\ref{A59}) show that step-like and
barrier-like potentials become transparent in the case of
the so-called "normal incidence", $k_y=0$ (or $\phi=\bar{\phi}=0$), which produces
the Klein paradox in graphene \cite{Kats}. However, such transparency of potentials does not occur if exact
tight-binding solutions are written in the low-energy limit. Indeed,
the expression for $\cos\theta$ is determined by
the general expansion (\ref{A49}),
\begin{equation}\label{A49a}
\cos\theta=-\frac{{\rm
sgn}(E)}{\sqrt{k_x^2+k_y^2}}\left[k_y+\frac{a_{\rm
C-C}}{4}\,\left(3k_x^2+k_y^2\right)+\ldots\,\right].
\end{equation}
At $k_y=0$, the first term in (\ref{A49a}) vanishes
and the expression for $\cos\theta$ is described by the next non-vanishing term $O(k_x,k_y)$,
\begin{equation}\label{A60}
\cos\theta=-\frac{3}{4}\,{\rm sgn}(E)k_xa_{\rm
C-C}=-\frac{E}{2},\qquad |E|\ll 1.
\end{equation}

Inclusion of the next expansion terms in the expressions for $\cos\theta$ and $\cos\bar{\theta}$ resembles extending the  standard $\bf{ k\cdot p}$ method by including higher order $\bf{ k\cdot p}$ terms (see Ref. \cite{Ando2} and Eqs. (2.2) and (4.1) therein). For instance, the effect of small backscattering in graphene structures  was predicted for the first time within the framework of the extended model \cite{Ando2}. Thus, we conclude that the relativistic approach based on the standard $\bf{ k\cdot p}$ scheme should not be applied without amendments for describing electron transport through the gapless modes. The next non-vanishing term of the expansion (\ref{A49a}) "disables" the Klein paradox: the expression $(\cos\theta-\cos\bar{\theta})^2$ entering Eqs. (\ref{A35}) and (\ref{A44}) becomes quadratic on $U_0$, which results in the same order deviation of the exact transmission coefficients (\ref{A35}) and (\ref{A44}) from  unity (see Eq. (\ref{A57})).

Evidently, since $|k_y|$ increases proportionally to $|k^m-2\pi/3|$, we can restrict attention only to the first term in (\ref{A49a}) if $|k_y|$ is large enough, which makes the analytical solutions (\ref{A35}) and (\ref{A44}) similar to the approximate dependencies (\ref{A58}) and
(\ref{A59}) for the Dirac formalism.

\section{Modeling and discussion}\label{11}

\begin{figure}
\centering
\includegraphics[width=\textwidth]{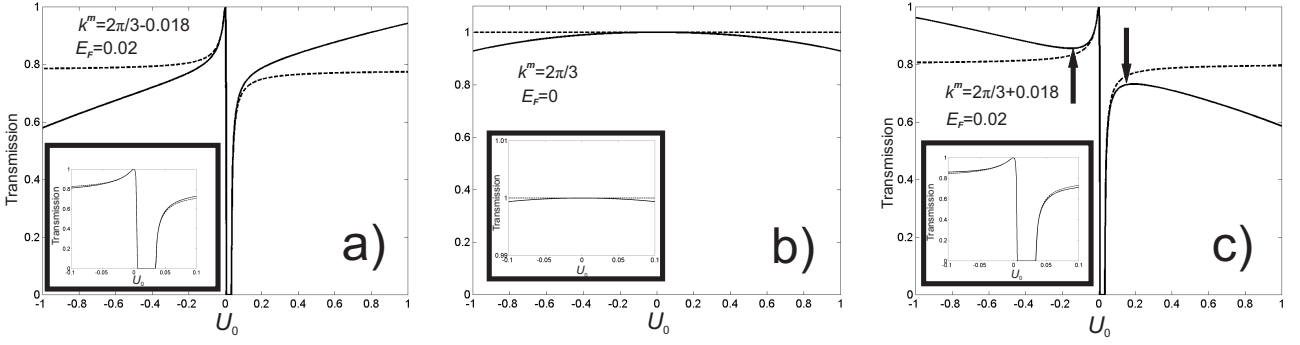}
\caption{Through-step transmission coefficients versus the applied
potential energy $U_0$ for three different transverse wave numbers
$k^m$: a) $k^m=2\pi/3-0.018$ and $E_F=0.02$; b) $k^m=2\pi/3$,
$E_F=0$; c) $k^m=2\pi/3+0.018$, $E_F=0.02$. Solid lines
correspond to the exact solution (\ref{A35}), dashed lines to the approximate
dependence (\ref{A58}). The inserts show the same curves in the
low-voltage region $|U_0|<0.1$. Two arrows point to the peak and dip of the exact transmission coefficient
(\ref{A35}).} \label{PRBstepU0}
\end{figure}

Now we compare the exact tight-binding solutions (\ref{A35}) and
(\ref{A44}) with the approximate expressions (\ref{A58}) and
(\ref{A59}). Figure \ref{PRBstepU0} represents
$U_0$-dependence of the through-step transmission coefficients
(\ref{A35}) and (\ref{A58}) for three different values of $k^m$:
(a) $k^m=2\pi/3-0.018$, (b) $k^m=2\pi/3$,  and (c) $k^m=2\pi/3+0.018$.
We assume the electron energy $E$ is fixed: (a,c) $E=E_F=0.02$
and (b) $E=E_F=0$. Simple estimations show that this choice leads to  single-mode propagation in graphene ribbons
having (a) ${\cal N}=58$, (b) ${\cal N}=59$, and (c) ${\cal N}=60$
elementary cells in the transverse direction. These curves
are very closed to each other when $|U_0|<0.1$, as depicted in
Fig. \ref{PRBstepU0}. Thus, the approximation (\ref{A58}) for the low-voltage region is numerically indistinguishable from the exact solution (\ref{A35}).

When $|U_0|<1$, the exact tight-binding solution
(\ref{A35}) for $k^m=2\pi/3+0.018$, unlike the approximate
solution (\ref{A58}), reveals two extreme points, as indicated
in the Figure \ref{PRBstepU0}c. The existence of such
transmission peculiarities follows from the definition (\ref{A36}) an the following chain of
identities
\[
\frac{dT}{dU_0}=\frac{dT}{d\bar{\theta}}\,\frac{d\bar{\theta}}{dU_0}=
\frac{dT}{d\cos\bar{\theta}}\,
\frac{d\cos\bar{\theta}}{dU_0}=\frac{dT}{d\cos\bar{\theta}}\,\frac{dg(E-U_0)}{dU_0}=
-\frac{dT}{d\cos\bar{\theta}}\,\left.\frac{dg(E)}{dE}\right|_{E\rightarrow E-U_0}.
\]
Thus, the additional peak and dip of the transmission
coefficient are the extreme points of $g(E,k^m>2\pi/3)$, $\pm\sqrt{1-4\cos^2(k^m/2)}$, as discussed in Sec.
\ref{9}.

It is also important that the exact dependence (\ref{A35}) for $k^m\ne 2\pi/3$ predicts unit transmission at
applied electrostatic energy $U^{res}_0=E_F-x/E_F$, which is the
solution to (\ref{A55}) with respect to $U_0$. However, in the
case under consideration these resonances are out of the indicated
region: $U^{res}_0>1$ for $k^m<2\pi/3$, and $U^{res}_0<-1$ if
$k^m>2\pi/3$.

\begin{figure}
\centering
\includegraphics[width=\textwidth]{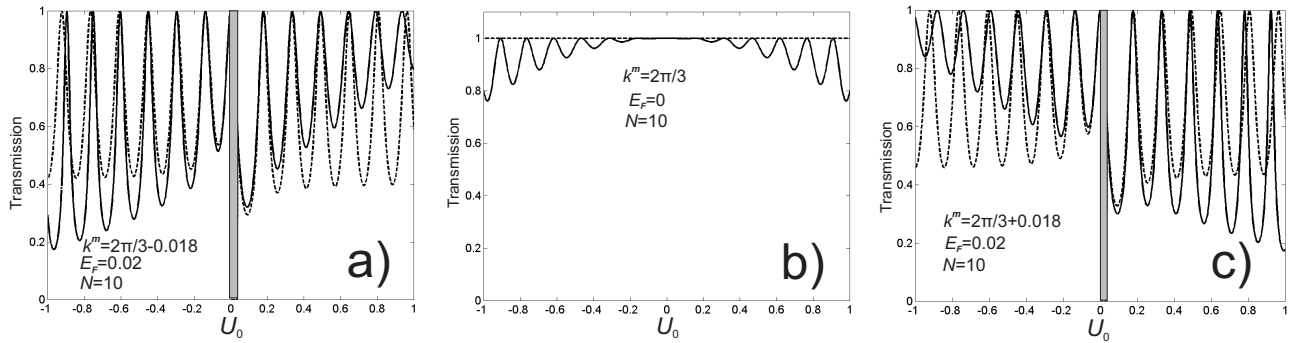}
\caption{Through-barrier transmission coefficients (\ref{A44})
(solid line) and (\ref{A59}) (dashed line)  versus $U_0$
calculated for a) $k^m=2\pi/3-0.018$ and $E_F=0.02$; b)
$k^m=2\pi/3$, $E_F=0$; c) $k^m=2\pi/3+0.018$, $E_F=0.02$. The grey
rectangle is the region of tunneling through a gap which
is not studied here.} \label{PRBbarU0n10}
\end{figure}

Figure \ref{PRBbarU0n10} depicts the through-barrier transmission
coefficients (\ref{A44}) and (\ref{A59}) for the same values of
$k^m$ and $E_F$, as Fig. \ref{PRBstepU0}.
To make the curves more distinguishable, we consider a
narrow barrier, $N=10$. One can
see the series of pronounced peaks associated with the
resonance conditions $\bar{k}^nN\equiv \bar{k}_xD=\pi l$
($l=0,1,\ldots$). A visual comparison of the approximation (\ref{A59})
with the exact solution (\ref{A44}) shows that they are closed to each other in the
low-voltage region. However, there are some discrepancies in the resonance
positions, especially in the depth of dips outside the low-voltage
region. Discrepancies of this kind are also observed between the exact and
approximate through-step solutions (\ref{A35}) and (\ref{A58}) over the
entire $U_0$-region. Formally, if the barrier-induced system is considered as a combination of two scatterers -- interfaces between
the external and disturbed leads, the through-barrier probability
$T^{\rm bar}$ can be written down in terms of the through-step
transmission $T^{\rm step}$. The probability $T^{\rm bar}$ varies within the interval
defined by (32) from Ref. \cite{Klimenko}
$$\left(\frac{T^{\rm step}}{2-T^{\rm step}}\right)^2\leq T^{\rm
bar}\leq 1,$$ which can also be observed from the behaviour of the
corresponding through-step coefficient depicted in Fig.
\ref{PRBstepU0}.

Obviously, as $|k^m-2\pi/3|$ increases the exact (tight-binding) and
approximate solutions for through-step and through-barrier
transmissions in the low-voltage region ($|U_0|\ll
1$) are closer to each other. This also follows from Sec.
\ref{10}, specifically from the expansion (\ref{A49a}).

Now we consider propagation through a gapless band
($k^m=2\pi/3$). Despite visual coincidence of the exact
tight-binding solutions with unity in the low-voltage region
(Figs. \ref{PRBstepU0}b and \ref{PRBbarU0n10}b) the dependences
(\ref{A35}) and (\ref{A44}) do not reveal unit propagation for the whole low-voltage
region. Thus, the Klein paradox, which strongly predicts that $T=1$ for one-mode propagation through any high and wide
potential barriers in semimetal ribbons, fails since the
through-step and through-barrier
coefficients $T^{{\rm step}({\rm bar})}$ differ from unity though the corresponding
deviations are indistinguishable
at experimentally reliable values $|U_0|\ll 1$.

\section{Conclusions}\label{12}

In the framework of the tight-binding model, we present a full and closed description of electron transport in armchair GRs.
Starting with the tight-binding Schr\"{o}dinger
equations for a 2D honeycomb lattice, we identify the boundary conditions, wave function and propagation velocity
of an electron in armchair GRs. We obtain analytical expressions for
the through-step and through-barrier transmission
coefficients and demonstrate that new type
transmission resonances exist. Such resonances occur when
$\cos\theta=\cos\bar{\theta}$, where $\theta$ and
$\bar{\theta}$ are the inter-cell electron phase shifts in the corresponding regions.
We show that the number and positions of the resonances are strongly dependent on the electron transverse wave number
$k^m$. In particular, at $k^m=2\pi/3$ (propagation through a gapless
band), the resonances are absent, and the through-step
transmission coefficient reduces to the relation known for through-step propagation in a linear chain of
identical atoms. For the low-energy limit, the deviation of
the through-step transmission coefficient from unity is proportional to the square of applied potential energy. A
similar deviation is also observed for the through-barrier
transmission coefficient.

The discrepancy between our tight-binding result and the Klein
paradox in graphene becomes evident as a result of expanding the expressions for
$\cos\theta$ and $\cos\bar{\theta}$ in the vicinity of the zero-energy
point. We show that in the low-energy limit these
expressions come to negligibly small terms, which are ignored in the well-known
$\bf{ k\cdot p}$ method leading to the relativistic Dirac equation.
These negligibly small terms "destroy" unit propagation and produce a small backscattering proportional to $U_0^2$.

The presented analytical results are in complete agreement with
the results of the numerical computations made in the paper.

\begin{acknowledgments}
The authors are deeply thankful to Pavlo Prokopovych for editing the manuscript. This paper has been supported by STCU Grant \# 21-4930/08.
\end{acknowledgments}

\appendix

\section{Computation of electron flux}\label{A}

In this appendix, we compute the flux related to electron wave propagation in armchair graphene ribbons. We
start with the set of non-stationary Schr\"{o}dinger equations
$$\begin{array}{ll}\left\{\!\!
\begin{array}{ll}
\dot{\psi}_{n,m,l}=-\frac{i\beta}{\hbar}\left[\psi_{n,m,\lambda}+\psi_{n,m+1,\lambda}+\psi_{n-1,m,r}\right],\\
\dot{\psi}_{n,m,\lambda}=-\frac{i\beta}{\hbar}\left[\psi_{n,m,l}+\psi_{n,m,\rho}+(1-\delta_{m,1})\psi_{n,m-1,l}\right],
\\
\dot{\psi}_{n,m,\rho}=-\frac{i\beta}{\hbar}\left[\psi_{n,m,\lambda}+\psi_{n,m,r}+(1-\delta_{m,1})\psi_{n,m-1,r}\right],
\\
\dot{\psi}_{n,m,r}=-\frac{i\beta}{\hbar}\left[\psi_{n,m,\rho}+\psi_{n,m+1,\rho}+\psi_{n+1,m,l}\right],
\end{array}
\right. & 1\leq m\leq {\cal N},\\ \left\{\!\!
\begin{array}{l}
\dot{\psi}_{n,{\cal
N}+1,\lambda}=-\frac{i\beta}{\hbar}\left(\psi_{n,{\cal
N}+1,\rho}+\psi_{n,{\cal N},l}\right),
\\
\dot{\psi}_{n,{\cal
N}+1,\rho}=-\frac{i\beta}{\hbar}\left(\psi_{n,{\cal
N}+1,\lambda}+\psi_{n,{\cal N},r}\right).
\end{array}
\right.&
\end{array}$$
Multiplying the equations for $\dot{\psi}_{n,m,\alpha}$ by
$\psi^{*}_{n,m,\alpha}$ ($m=1,\ldots,{\cal
N}+1$) and adding the obtained relations to the complex conjugated counterparts, and then summing them up over all carbon atoms
in unit cell $n$, we get the following
$$ \frac{d}{dt}\left[\sum_{m=1}^{{\cal
N}}\left\{\left|\psi_{n,m,l}\right|^2+\left|\psi_{n,m,r}\right|^2\right\}+\sum_{m=1}^{{\cal
N}+1}\left\{\left|\psi_{n,m,\lambda}\right|^2+\left|\psi_{n,m,\rho}\right|^2\right\}\right]=
J^{left}+J^{right}, \eqno{(A1)}$$ where $$
J^{left}=-\frac{i\beta}{\hbar}\sum_{m=1}^{{\cal
N}}\left[\psi_{n-1,m,r}\psi^{*}_{n,m,l}-\psi^{*}_{n-1,m,r}\psi_{n,m,l}\right],
$$ and
$$ J^{right}=-\frac{i\beta}{\hbar}\sum_{m=1}^{{\cal
N}}\left[\psi_{n+1,m,l}\psi^{*}_{n,m,r}-\psi^{*}_{n+1,m,l}\psi_{n,m,r}\right].
$$ The left side of Eq.(A1) is the rate of change of the electron
probability in unit cell $n$. The right side of the equation
gives the flux entering to the unit cell. Thus, the flux $J$
outgoing from unit cell $n$ to the right direction equals $$
J=-J^{right}=\frac{i\beta}{\hbar}\sum_{m=1}^{{\cal
N}}\left[\psi_{n+1,m,l}\psi^{*}_{n,m,r}-\psi^{*}_{n+1,m,l}\psi_{n,m,r}\right].\eqno{(A2)}
$$ This formula is used in the main text.

\section{Through-step transmission in a linear chain of atoms}\label{B}

\begin{figure}
\centering
\includegraphics[width=0.32\textwidth]{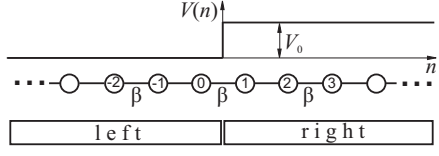}
\caption{Model of the linear chain of atoms coupled by the
electron-transfer integral $\beta$. As in the main text, atoms
with $n\leq 0$ correspond to the left lead, and $n \geq 1$ -- to
the right lead.}\label{LinChainAtoms}
\end{figure}

In this section, we calculate the through-step transmission coefficient
for electron moving in a linear chain of identical atoms (see Fig.
\ref{LinChainAtoms}). For the sake of convenience, we keep the same notation
for $E\equiv E/|\beta|$ and $U_0=eV_0/|\beta|$.

Following Fig. \ref{LinChainAtoms}, we need to determine the solution to the following system of linear
Schr\"{o}dinger equations
$$
\left\{\!\!
\begin{array}{l}
-E\psi^{^{\rm left}}_{n}=\psi^{^{\rm left}}_{n-1}+\psi^{^{\rm
left}}_{n+1}, \quad n<0,\\ -E\psi^{^{\rm left}}_{0}=\psi^{^{\rm
left}}_{-1}+\psi^{^{\rm right}}_{1}, \\ (U_{0}-E)\psi^{^{\rm
right}}_{1}=\psi^{^{\rm left}}_{0}+\psi^{^{\rm right}}_{2},\\
(U_{0}-E)\psi^{^{\rm right}}_{n}=\psi^{^{\rm
right}}_{n-1}+\psi^{^{\rm right}}_{n+1},\quad n>1,
\end{array}
\right.\eqno{({\rm B}1)}
$$
where
$$
\psi^{^{\rm left}}_{n}=e^{ik n}+re^{-ik n},\qquad \psi^{^{\rm
right}}_{n}=te^{i\bar{k} n},\eqno{({\rm B}2)}
$$
$r$ and $t$ are the amplitudes of the reflected and transmitted
waves, respectively, $k$ and $\bar{k}$ are the non-dimensional wave numbers in the
left and right regions. One can see that the first and forth
equations in (B1) are satisfied identically if
$$
\cos k=-\frac{E}{2},\qquad \cos
\bar{k}=-\frac{E-U_0}{2}.\eqno{({\rm B}3)}.
$$
Plugging in (B2) into the second and third equations from (B1) gives the
system of two linear equations $$1+r=t,\qquad
e^{ik}+e^{-ik}\,r=e^{i\bar{k}}\,t.$$ The solution to the system is
$$
r=-\frac{e^{i\bar{k}}-e^{ik}}{e^{i\bar{k}}-e^{-ik}},\qquad
t=\frac{2i\sin k}{e^{i\bar{k}}-e^{-ik}}.
$$
Thus, the transmission coefficient determined as $T=|t|^2\sin
\bar{k}/\sin k$ can be represented as
$$
T=\frac{4\sin k\,\sin\bar{k}} {4\sin k
\,\sin\bar{k}+\left(\sin\bar{k}- \sin
k\right)^2+\left(\cos\bar{k}-\cos k \right)^2}.\eqno{({\rm B}4)}
$$
Since the values of $\sin k$ and $\sin\bar{k}$ are positive in the
corresponding bands ($0\leq (k,\bar{k})\leq \pi$) one can come to an expression similar to (\ref{A57a}) for the transmission coefficient.

Therefore, independently on the Fermi position inside the band,  electron propagation through a
potential step in a graphene ribbon is described by the same
through-step relation as in an undimerized 1D chain with
the same ratio $U_0=eV_0/|\beta|$.

\section{Relativistic Dirac solution for the electron reflection probability through a potential barrier.}\label{C}

The reflection amplitude from a potential barrier obtained through the Dirac formalism is described by Eq. (3) in Ref. \cite{Kats}
$$
r=\frac{2ie^{i\phi}\sin
(\bar{k}_xD)\,[\sin\phi-ss'\sin\bar{\phi}]}{ss'\left[
e^{-i\bar{k}_xD}\cos(\phi+\bar{\phi})+e^{i\bar{k}_xD}\cos(\phi-\bar{\phi})\right]-2i\sin
(\bar{k}_xD)},\eqno{({\rm C}1)}
$$
where
$s={\rm
sgn}(E)$ and $s'={\rm
sgn}(E-U_0)$.
Equivalently,
$$
r=\frac{ie^{i\phi}\sin
(\bar{k}_xD)\,[\sin\phi-ss'\sin\bar{\phi}]}
{ss'\cos(\bar{k}_xD)\cos\phi\cos\bar{\phi}+i\sin(\bar{k}_xD)[ss'\sin\phi\sin\bar{\phi}-1]}.
$$
Since
$$(1-ss'\sin\phi\sin\bar{\phi})^2=(\sin\phi-ss'\sin\bar{\phi})^2+\cos^2\phi\cos^2\bar{\phi},$$
the reflection probability $R=|r|^2$ is given by
$$
R=\frac{\sin^2
(\bar{k}_xD)\,[\sin\phi-ss'\sin\bar{\phi}]^2}
{\cos^2\phi\cos^2\bar{\phi}+\sin^2(\bar{k}_xD)[\sin\phi-ss'\sin\bar{\phi}]^2}.
$$
Thus, the transmission probability $T=1-R$ through the potential barrier is exactly the same as our approximate relation (\ref{A59}).

\end{document}